\documentclass[twocolumn,trackchanges,twocolappendix]{aastex63}
\usepackage{lineno}
\usepackage{amsmath}
\usepackage{graphicx}
\usepackage{epstopdf}
\usepackage{color}
\definecolor{marine}{RGB}{255,0,127}

\received{February 29, 2021}
\revised{February 30, 2022}
\accepted{February 31, 2023}
\submitjournal{ApJ}

\shorttitle{A Red Giants' Toy Story II}
\shortauthors{Miller Bertolami, M. M.}
\begin{document}

\title{A Red Giants' Toy Story II: Understanding the Red-Giant Branch Bump}

\correspondingauthor{Marcelo M. Miller Bertolami}
\email{mmiller@fcaglp.unlp.edu.ar, marcelo@mpa-garching.mpg.de}

\author[0000-0001-8031-1957]{Marcelo M. Miller Bertolami}
\affiliation{Instituto de Astrof\'isica de La Plata, Consejo Nacional de Investigaciones Cient\'ificas y T\'ecnicas \\
Avenida Centenario (Paseo del Bosque) S/N,  B1900FWA La Plata, Argentina.}

\affiliation{Facultad de Ciencias Astron\'omicas y Geof\'isicas,  Universidad Nacional de La Plata\\
  Avenida Centenario (Paseo del Bosque) S/N,  B1900FWA La Plata, Argentina.}



\begin{abstract}

  The Red-Giant Branch Bump (RGBB) is one of the most noteworthy
  features in the red-giant luminosity function of stellar
  clusters. It is caused by the passage of the hydrogen-burning shell
  through the composition discontinuity left at the point of the deepest
  penetration by the convective envelope. When crossing
  the discontinuity the usual trend in increasing luminosity reverses
  for a short time before it increases again, causing a zig- zag in
  the evolutionary track.

  In spite of its apparent simplicity the actual physical reason
  behind the decrease in luminosity is not well understood and
  several different explanations have been offered.

 Here we use a recently proposed simple toy-model for the
  structure of low-mass red giants, together with previous results, to
  show beyond reasonable doubt that the change in luminosity at the
  RGBB can be traced to the change in the mean molecular weight of the
  layers on top of the burning shell. And that these changes happen on
  a nuclear timescale. The change in the effective mean molecular
  weight, as the burning shell approaches the discontinuity, causes a
  drop in the temperature of the burning shell which is attenuated by
  the consequent feedback contraction of the layers immediately below
  the burning shell.

   Our work shows that, when applied correctly, including the feedback
    on the structure of the core together with of the increase in the
    mass of the core, shell-source homology relations do a great
    quantitative job in explaining the properties of full evolutionary
    models at the RGBB.
\end{abstract}



\section{Introduction} \label{sec:intro}
\begin{figure}
\centering
\includegraphics[width=\columnwidth]{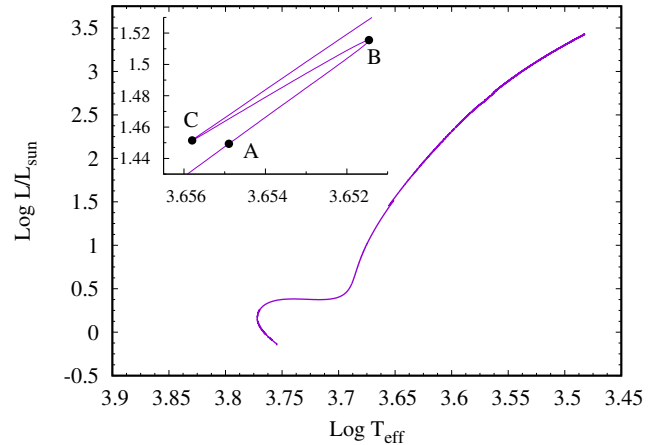}
\caption{Evolutionary track of a $1 M_\odot$ model (initial
  composition $X=0.695$ and $X=0.02$). Inset: Zoom in the region of
  the RGBB. Letters A, B and C indicate the location of the models
  snapshots discussed in the text.}
\label{fig:HR}
\end{figure}

It was discovered already in the early days of automatic stellar
evolution computations that low-mass stars undergo a brief phase of
decreasing luminosity \citep{1967ZA.....67..420T, 1968ApJ...154..581I}
during the red giant branch (RGB). This drop in luminosity is caused
by the passage of the hydrogen(H)-burning shell through the
composition discontinuity left by the deepest penetration of
convection into the stellar envelope (Fig \ref{fig:HR}). This creates
a zig-zag in the evolutionary track, and the star crosses the same
luminosity region three times. Consequently, red giants spend a little
longer in that region of the Hertzsprung–Russell (HR) diagram. In
stellar clusters this phase corresponds to an accumulation of stars at
that specific luminosity. This produces a bump in the luminosity
function of red giant branch stars that was first measured by
\cite{1985ApJ...299..674K}, and it is usually known as the red-giant
branch bump (RGBB).

  In spite of its apparent simplicity the actual physical mechanism
  behind the sudden decrease in luminosity is not well
  understood. \cite{1968ApJ...154..581I} suggested that the drop in
  the stellar luminosity was a direct effect of the increase in the
  abundance of H in the burning shell when crossing the chemical
  discontinuity. Despite the absence of a clear mechanism for this
  connection several other authors have concurred with this position
  \citep[e.g.][]{2002ApJ...565.1231C,2015ApJ...804....6G}.
  Nevertheless, a detailed analysis of stellar models around the RGBB
  by \cite{1990ApJ...364..527S} showed that luminosity starts dropping
  before the hydrogen-burning shell actually reaches the hydrogen
  discontinuity. They concluded that the luminosity drop cannot be due
  to the burning shell responding to the increase in the available
  fuel, as that fuel has not yet been reached. Instead, they suggested
  that the drop in luminosity should be due to the increase in the
  opacity above the burning shell that results from the higher H
  abundance. Taking a completely different approach
  \cite{2020MNRAS.492.5940H} analyzed the temporal changes in the
  entropy distribution during the drop in luminosity at the RGBB.  A
  more likely explanation of the RGBB was suggested by
  \cite{1970A&A.....6..426R} who, under the assumption of the so
  called shell-source homology relations (see appendix
  \ref{app:shellhomology}), noticed that the drop in the mean
  molecular weight ($\mu$) at the transition should cause a drop in
  the luminosity of the burning shell. This idea was further explored
  by \cite{2015MNRAS.453..666C} who studied in detail the impact of
  variations of $\mu$ in the layers immediately above the burning
  shell.  \cite{2015MNRAS.453..666C} concluded that it is plausible
  that the mean molecular weight above the burning shell is the main
  cause of the drop in luminosity. The main problem with this
  explanation, as noted by \cite{2015MNRAS.453..666C}, is the
  substantial departure in the predictions of shell-source homology
  relations from those of full evolutionary models (FEMs), which calls
  into question the validity of the argument.

  Recently, we have developed a simple solution the long standing
  question of why stars become red giants \citep{2022-ToyStory}. As
  part of this explanation we devised a quantitative toy-model for
  low-mass red giants (Fig. \ref{fig:Structure}). One of the key
  insights from this model is that the location of the burning shell
  ($R_s$) is not independent from the temperature of the shell
  ($T_s$). Consequently, when the burning shell approaches the
  chemical discontinuity the decrease in $\mu$ immediately above the
  burning shell leads to a drop in temperature in the isothermal layer
  between the degenerate core and the burning shell (see Fig
  \ref{fig:Structure}). As this isothermal region has the equation of
  state of an ideal classical gas, this cooling leads to a
  contraction. According to shell-source homology relations, this feedback of
  the temperature drop on the location of the burning shell should
  also affect the luminosity of the burning shell. In this paper we
  show how the feedback of the radius of the burning shell leads to a
  very good agreement between the predictions of the simple model and
  those of FEMs, proving beyond reasonable doubt that it is the change
  in $\mu$ what causes the RGBB.

%%%%%%%%%%%%%%%%%%%%%%%%%%%%%%%%%%%%%%%%%%%%%%%%%%%%%%%%%%%%%%%%%%%5

%\section{A simple red giant model}
\section{Definitions, shell-source homology relations and simple models}
\label{sec:toy-model}
\begin{figure}
\centering \includegraphics[width=\columnwidth]{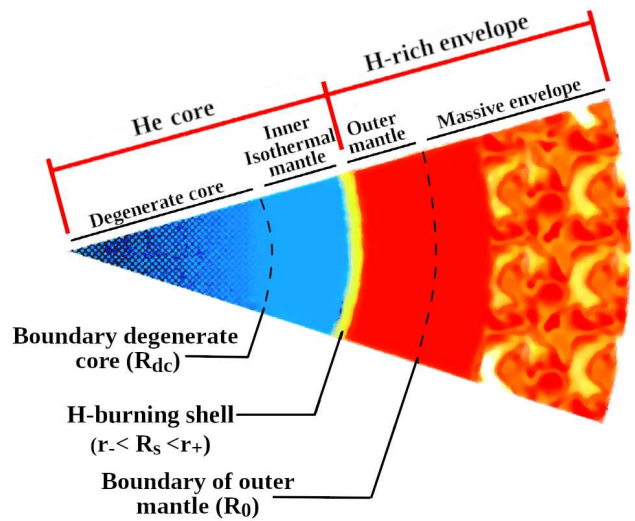}
\caption{Main structural parts and definitions of low-mass red giants
  and the toy model developed in \cite{2022-ToyStory}. The inner and outer  mantles around the burning shell are, in most cases, nondegenerate and massless. }
\label{fig:Structure}
\end{figure}

For the sake of clarity we will first define some relevant
  quantities. Following \cite{2022-ToyStory} the structure of a
  low-mass red giant can be described as consisting of a degenerate
  core of mass ($M_c$) and radius ($R_{\rm dc}$) surrounded by a
  massless isothermal mantle where the gas behaves as an ideal
  classical gas. Above sits the H-burning shell, where heat is being
  released by nuclear burning. We define $r_{-}$ and $r_{+}$ as the
  lower and upper boundaries of the burning shell, where the local
  luminosity ($l(r)$) goes from the value at the core ($L_c$) to the
  surface value $L_\star=L_c+L_s$, where $L_s$ is the total power
  released by the burning shell. Due to degeneracy of the electron
  gas, the core contracts only due to the increase in its mass, which
  happens on a nuclear timescale making the heat released by
  gravitational contraction $L_c\ll L_s \simeq L_\star$. We define the
  nominal location of the burning shell ($R_s$) as the point at which
  $l(R_s)\simeq L_s/2$, which is very close to the maximum in the
  energy generation rate. Thanks to the high temperature sensitivity
  of nuclear reactions the burning shell can be assumed be thin,
  $|r_{+} - r_{-}|\ll R_s$. Under this assumption $R_s$ is also the
  radius of the helium (He) core. Above the burning shell it is useful
  to define an outer mantle between $r_+$ and an arbitrary point $R_0$
  at which pressure, temperature and density have dropped by orders of
  magnitude from their values at the burning shell ($P_0 \ll P_s$,
  $T_0 \ll T_s$, $\rho_0 \ll \rho_s$). Let us note that above $r_+$
  already $l(r)=L_\star\simeq L_s$ and the composition is that of the
  hydrogen rich envelope (corresponding to a mean molecular weight
  $\mu_{\rm env}$). As discussed in \cite{2022-ToyStory} as soon as
  the core is dense ($\rho_s\ll 4\pi M_c / 3 {R_s}^3 $) the outer
  mantle can be considered massless $\Delta m\ll M_c$.

 One usual way of understanding the behavior of burning shells is with
the help of the so-called shell-source homology relations developed by
\cite[][]{1970A&A.....6..426R}. Shell-source homology rests on several
assumptions (see appendix \ref{app:shellhomology}), in particular it
is assumed that the region between $r_{-}$ and $R_0$ is massless
($\Delta m \ll M_c$), and that the solutions $\rho(r)$, $T(r)$,
$P(r)$, and $l(r)$ of the stellar structure equations only depend on
$M_c$, $R_s$ and the mean molecular weight ($\mu$) through simple
power laws. Under these assumptions it is possible to show that two
different solutions $\rho(r)$, $T(r)$, $P(r)$, $l(r)$ (corresponding
to $M_c$,$R_s$, and $\mu$), and $\rho'(r')$ $T'(r')$, $P'(r')$,
$l'(r')$ (corresponding to $M_c'$,$R_s'$, and $\mu'$) evaluated at
homologous points ($r/R_s=r'/R_s'$) are related by
%\begin{equation}
  \begin{align}
    \frac{\rho}{\rho'}=& \left(\frac{M_c}{M_c'}\right)^{(4-\nu)/3} \left(\frac{R_s}{R_s'}\right)^{(-6+\nu)/3}\left(\frac{\mu}{\mu'}\right)^{(4-\nu)/3} \label{eq:sh-thomson-cno_1}\\
    \frac{T}{T'}=&\left(\frac{M_c}{M_c'}\right) \left(\frac{R_s}{R_s'}\right)^{-1}\left(\frac{\mu}{\mu'}\right) \label{eq:sh-thomson-cno_2}\\
    \frac{P}{P'}=& \left(\frac{M_c}{M_c'}\right)^{(7-\nu)/3} \left(\frac{R_s}{R_s'}\right)^{(-9+\nu)/3}\left(\frac{\mu}{\mu'}\right)^{(4-\nu)/3} \label{eq:sh-thomson-cno_3}\\
    \frac{l}{l'}=& \left(\frac{M_c}{M_c'}\right)^{(8+\nu)/3} \left(\frac{R_s}{R_s'}\right)^{(-3-\nu)/3}\left(\frac{\mu}{\mu'}\right)^{(8+\nu)/3},
    \label{eq:sh-thomson-cno_4}
 \end{align}
%\end{equation}  
where we have used a Thomson scattering opacity ($\kappa=\kappa_0 T^a
P^b=\kappa_0$, i.e. $a=b=0$) and a typical CNO energy generation rate
($\epsilon=\epsilon_0 \rho T^\nu$).  It is also worth noting that, as
$\mu(r)$ changes in the region of the burning shell, to obtain
eqs. \ref{eq:sh-thomson-cno_1} to \ref{eq:sh-thomson-cno_4} one needs
to assume that $\mu'(r'/R_s')$ can be obtained by scaling up the
function $\mu(r/R_s)$ by the same factor at each homologous
point. This is clearly not completely accurate, as the value at the
bottom of the burning shell ($r=r_{-}$) in FEMs has to remain constant
and equal to the mean molecular weight of the core
($\mu_{-}=\mu_{-}'=\mu_c$). Alternatively, the values of $\mu$ and
$\mu'$ can be understood, within the framework of shell-source
homology relations as a proper average of the mean molecular weight in
relevant region \citep[see][ for a discussion]{1970A&A.....6..426R}.

Interestingly in \cite{2022-ToyStory} we have shown that, when
  the core is dense enough it is possible to prove that the values of
  $\rho_s$, $T_s$, $P_s$ in the middle of the burning shell, and the
  total luminosity $L_s$ of the burning shell are only dependent on
  the values of $M_c$ and $R_s$ as assumed in shell-source homology
  relations. Moreover, it is possible to show that these quantities
  fulfill relationships similar to eqs. \ref{eq:sh-thomson-cno_1} to \ref{eq:sh-thomson-cno_4}.  Under
  the assumption of an inert core ($L_c=0$), Thomson scattering, and a
  typical CNO energy generation rate eqs. 32, 35, 36 and 37 of \cite{2022-ToyStory} tell us that
%  \begin{equation}
  \begin{align}
    \rho_s=& \mathcal{K} {M_c}^{(4-\nu)/3} {R_s}^{(-6+\nu)/3} \mu_{\rm env}^{(2-\nu)/3} {\mu_s}^{2/3}  \label{eq:mb22-thomson-cno_1}\\
    T_s=& \mathcal{K'}{M_c}{R_s}^{-1} {\mu_{\rm env}} \label{eq:mb22-thomson-cno_2}\\
    P_s=& \mathcal{K''}{M_c}^{(7-\nu)/3} {R_s}^{(-9+\nu)/3}{\mu_{\rm env}}^{(5-\nu)/3}{\mu_s}^{-1/3} \label{eq:mb22-thomson-cno_3}\\
    L_s=&  \mathcal{K'''}{M_c}^{(8+\nu)/3} {R_s}^{(-3-\nu)/3}{\mu_{\rm env}}^{(7+\nu)/3}{\mu_s}^{1/3},
    \label{eq:mb22-thomson-cno_4}
 \end{align}
%\end{equation}  
where $\mathcal{K}$, $\mathcal{K'}$ , $\mathcal{K''}$,
$\mathcal{K'''}$ are constants, $\mu_s$ is the mean molecular weight
at the middle of the burning shell and $\mu_{\rm env}$ is the mean
molecular weight of the envelope (assumed to be constant in the outer
mantle). By comparing eqs. \ref{eq:sh-thomson-cno_1} to
\ref{eq:sh-thomson-cno_4} and eqs.  \ref{eq:mb22-thomson-cno_1} to
\ref{eq:mb22-thomson-cno_4} we see that under the assumption performed
in the derivation of shell-source homology relations, that changes in
$\mu_s$ are proportional to changes in $\mu_{\rm env}$, the two sets
of relations are formally similar. It is worth noting that, under the
framework provided by \cite{2022-ToyStory} the quantities in
eqs. \ref{eq:mb22-thomson-cno_1} to \ref{eq:mb22-thomson-cno_4}
correspond to the values of $\rho(r)$, $T(r)$, $P(r)$, $l(r)$ at
specific points and the meaning of the mean molecular weights are now
well defined. Due to the large value of $\nu$ it is clear from eqs.
\ref{eq:mb22-thomson-cno_1} to \ref{eq:mb22-thomson-cno_4} that it is
$\mu_{\rm env}$ what dominates the impact of changes in the molecular
weight on the burning shell. This should not be a surprise, as it is
only $\mu_{\rm env}$ what links $T_s$ to $M_c$ and $R_s$ and nuclear
burning is extremely sensitive to temperature. For the sake of
clarity, in the following discussion we will assume that $\Delta
\mu_s/\mu_s =\Delta \mu_{\rm env}/\mu_{\rm env}$ as usually done in
shell-source homology relations.  In real stars it is expectable that
relative changes in $\mu_s$ will be between $\Delta \mu_{\rm
  env}/\mu_{\rm env}$ and $\Delta \mu_{-}/\mu_{-}=0$. Interestingly,
due to the large value of $\nu$ this will be only a minor correction.

 \cite{2022-ToyStory} showed that  the core of a low-mass red
giant can be considered to good approximation as composed of two
parts, a degenerate core of mass $M\simeq M_c$ and radius $R_{\rm
  dc}\simeq 1.12\times 10^{20} {M_{\rm dc}}^{-1/3}.$ and an inner
nondegenerate isothermal mantle of negligible mass above (see
Fig. \ref{fig:Structure}). It is possible to show that the temperature, density, pressure and radius of the burning shell, and the mass of the core and mean molecular weight at and above the burning shell are not independent but
must fulfill
\begin{equation}
T_9 \simeq \frac{\mu_{\rm env}}{\mu_c}0.6 \frac{R_{\rm dc}}{R_s}
  \left(\frac{M_{\rm c}}{M_\odot}\right)^{4/3},
  \label{eq:1}
\end{equation}
and
\begin{equation}
  \begin{aligned} 
 1.633\times 10^{-26} &\frac{\mu_c^3}{\mu_s\mu_{\rm env}^2}\nu \exp(21/\nu)\left(1+1/\nu\right)^{-9/2} \\ = &{T_9}^{-1/6} \left[\frac{R_s}{R_{\rm dc}}\right]^2\left[\frac{M_c}{M_\odot}\right]^{-2/3} \\ 
 \times & \exp\left[\frac{12}{(1+2/\nu)}\frac{\mu_c}{\mu_{\rm env}}\left(1-\frac{R_{\rm s}}{R_{\rm dc}}\right) \right. \\
   -&\left.15.231 {T_9}^{-1/3}\right]
 \end{aligned} 
 \label{eq:2} 
\end{equation}  
where $T_9=T_s /10^9$ as it is common practice, and $R_{\rm dc}$
  is the radius of the degenerate part of the He core (see
  Fig. \ref{fig:Structure}).  Eqs. \ref{eq:1} and \ref{eq:2} have
been derived for the CNO-cycle and Thomson scattering opacities.

\section{Understanding the RGBB}

\begin{table*}[!]
  \begin{tabular}{l|cccccccc}
Snapshot  &  $M_c/M_\odot$	& $L_\star/L_\odot$ &     $R_s$ [cm] &	$R_{\rm dis}$ [cm]&  $T_s$[$10^7$ K]& $\rho_s$[g cm$^{-3}$] & $\mu_{+}$ &	$\mu_{\rm eff}$ \\\hline
model A  &  0.226146	& 28.1423 &   $2.1900\times10^9$ & $2.1164\times10^{10}$ & 2.8515 & 149.54 & 0.7212 & 0.7120 \\
model B  &  0.235181	& 32.7756 &   $2.1477\times10^9$ & $7.2146\times10^{9}$ & 2.9055 & 144.58 & 0.7124 & 0.6885 \\
model C  &  0.241617	& 28.2800 &   $2.0636\times10^9$ & - & 2.8741 & 158.08 & 0.6322 & 0.6322 \\
  \end{tabular}
  \caption{Properties of the structure of the stellar model at the
    snapshots displayed in Fig. \ref{fig:McL} ($1 M_\odot$ ($Z=0.02$)
    sequence). The value $\mu_{+}$ is taken immediately
    above the burning shell ($r\simeq r_{+}$).}
  \label{tab:snapshots}
\end{table*}

%\subsection{A naive implementarion of shell-homology relations}

Table \ref{tab:snapshots} shows the characteristics of the stellar
models before and after the RGBB (Figs. \ref{fig:McL} and
\ref{fig:r_m_H}). One of the characteristics of the models near
  the RGBB is the presence of the chemical discontinuity at $m_{\rm
    dis}\simeq 0.242 M_\odot$ ($r=R_{\rm dis}$) which was left by the
  maximum penetration of the convective envelope at earlier
  evolutionary stages.  From Table \ref{tab:snapshots} we see that
the relative drop in luminosity at the RGBB (from B to C) for our
reference model is $\delta L_\star/L_\star=(L_C-L_B)/L_B=-0.1372$,
while the relative drop in the mean molecular weight\footnote{Here the
  value of the mean molecular weight is taken immediately above the
  burning shell ($r\simeq r_{+}$), where the energy generation rate
  falls two orders of magnitude below the peak value.} is
$(\mu_C-\mu_B)/\mu_B=-0.1126$.  It is clear that the drop in
luminosity is much lower than what would be predicted from a naive use
of the shell-source homology relations
\cite[e.g.][]{1970A&A.....6..426R, 2012sse..book.....K}. For Thomson
scattering, which is a good approximation of the opacity at the
burning shell, and the typical assumption of $\nu=13$
\citep{2012sse..book.....K}\footnote{As we mention in appendix
  \ref{app:shellhomology} at the typical temperatures of the RGBB
  ($T_s\simeq 2.85 \times 10^7$K) the temperature dependence of the
  CNO cycle is closer to $\nu=16$. For the sake of comparison with
  previous works we keep $\nu=13$ in this estimations.}  the
dependence of the luminosity on the mean molecular weight predicted by
shell-source homology relations (eqs. \ref{eq:sh-thomson-cno_1} to \ref{eq:sh-thomson-cno_4}) is
$L_s\propto \mu^7$. For the change in $\mu_{+}$ in the outer mantle
between model B and C (see Table \ref{tab:snapshots}), this predicts
$\delta L_s/L_s =7\times \delta \mu/\mu= -0.7882$. We see that a naive
use of the shell-source homology relations manages to predict the
right trend in luminosity but errs by more than a factor 5.7. The
difference is still unacceptable if we choose the value of $\mu$
immediately below the discontinuity, $\mu^B_{\rm dis^{-}}=0.7066$, which gives $\delta
\mu/\mu=-0.1053$ and $\delta L_s/L_s =-0.7371$ (a factor 5.4 larger
than observed in the models).

 We will show below the agreement is improved by more than an order of
 magnitude when shell-source homology relations are applied in a more
 nuanced way.

\subsection{An improvement: the effective mean molecular weight of the outer mantle}
\label{sec:jcd}

\begin{figure}
\centering
\includegraphics[width=\columnwidth]{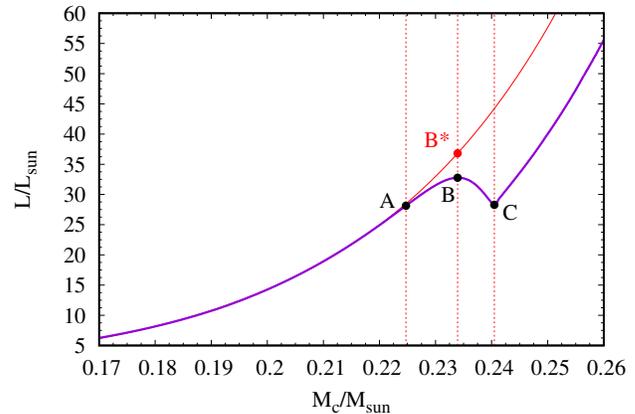}
\caption{The purple line shows the evolution of the luminosity as a
  function of the mass of the core for the FEM shown in
  Fig. \ref{fig:HR}. The red line shows an extrapolation of the
    evolution of $L(M_c)$ before point A, where the effect of the
    chemical discontinuity on the outer mantle is still not
    important. The difference between the red and purple lines is,
    then, a mesure of the impact of the chemical discontinuity on
    $L(M_c)$.}
\label{fig:McL}
\end{figure}

\begin{figure}
\centering
\includegraphics[width=\columnwidth]{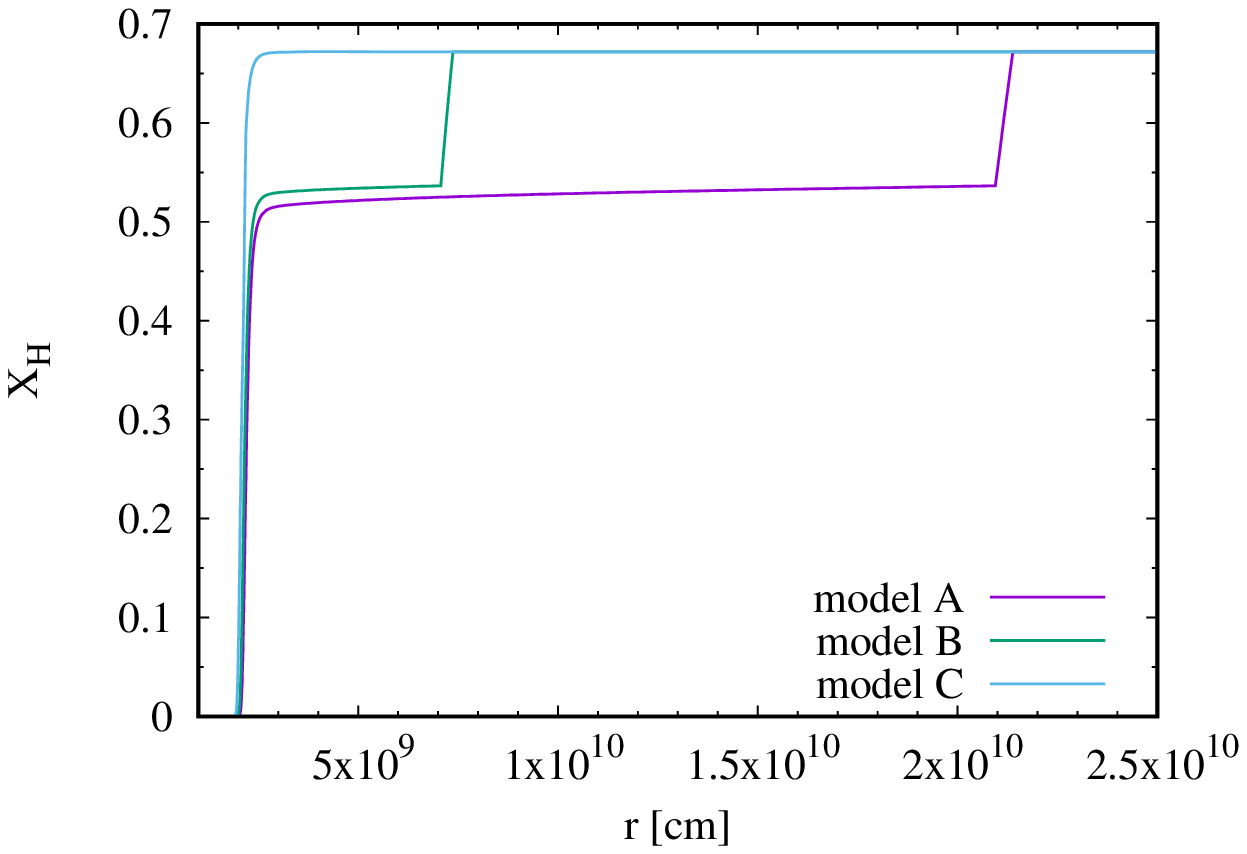}\\
\includegraphics[width=\columnwidth]{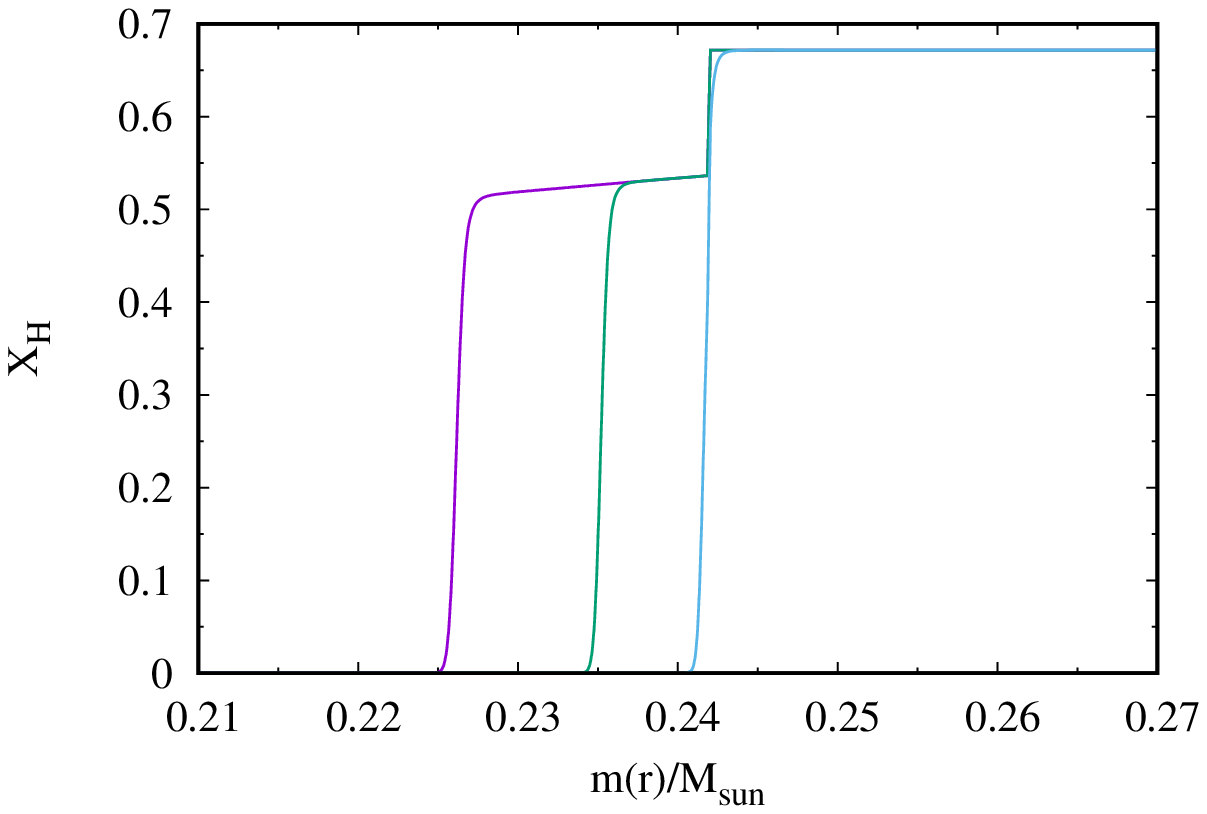}
\caption{Radial chemical profiles of the FEMs
  at different stages near the RGBB (see Figs. \ref{fig:HR} and
  \ref{fig:McL}). The chemical discontinuity left by the maximum
    penetration of the convective envelope at earlier evolutionary
    stages can be clearly appreciated at $m(r)\simeq 0.242
    M_\odot$.}
\label{fig:r_m_H}
\end{figure}

\cite{2015MNRAS.453..666C} studied in detail the impact of variations
of $\mu$ in the layers immediately above the burning
shell. \cite{2015MNRAS.453..666C} noted that, when the burning shell
approaches the discontinuity in the H profile, the relevant molecular
weight ($\mu_{\rm eff}$) that links $T_s$, $R_s$ and $M_c$,
\begin{equation}
   T_s\simeq \frac{G\,M_c \mu_{\rm eff}}{R_s\, \Re} \nabla_+,
\label{eq:T_MR}
\end{equation}
 is a mixture of the molecular weight above ($\mu_\uparrow$) and below
 ($\mu_\downarrow$) the discontinuity ($\nabla_+=1/4$ for Thomson
 scattering).  He showed that the effective molecular weight felt by
the burning shell can be written as
\begin{equation}
\mu_{\rm eff}=\mu_\downarrow\left[1-\frac{R_s}{R_{\rm dis}}\left(1-\frac{\mu_\uparrow}{\mu_\downarrow}\right)\right].
\label{eq:mueff}
\end{equation}  
  Eq. \ref{eq:T_MR} is, in fact, eq. \ref{eq:mb22-thomson-cno_2} and
  it is the reason why $\mu_{\rm env}$ appears in all the other
  equalities. This implies that, as the burning shell approaches the
  discontinuity in the chemical profile, it is $\mu_{\rm eff}$ what
  plays the role of $\mu_{\rm env}$ in
  eqs. \ref{eq:mb22-thomson-cno_1} to \ref{eq:mb22-thomson-cno_4}.
  Note that Eq. \ref{eq:mueff} was derived for a idealized situation
  where the mean molecular weight is strictly constant between the
  burning shell and the discontinuity which is not the case in real
  stars (Fig.\ref{fig:r_m_H}). When estimating $\mu_{\rm eff}$ in FEMs
  we use $\mu_\downarrow\simeq \mu_{+}$.  Using these expressions, we
  see that the mean molecular weight actually felt by the burning
  shell before reaching the discontinuity is slightly lower. The
  values for each snapshot are shown in the last column of Table
  \ref{tab:snapshots}.  With the corrected values we now see that,
  from B to C, the drop in the effective mean molecular weight is
  $\delta \mu_{\rm eff}/\mu_{\rm eff}=-0.0818$. Again, with the
  assumption of Thomson scattering and $\nu=13$, we have that
  $L_s\propto \mu_{\rm eff}^7$, this would imply $\delta L/L
  =-0.5724$, which is still a factor 4.2 higher than the actual
  value. As noted by \cite{2015MNRAS.453..666C} while this correction
  to the mean molecular weight improves the agreement with the drop in
  luminosity observed in FEMs, the disagreement is still
  substantial. This disagreement suggest some missing ingredients.

  Interestingly, one of the missing ingredients becomes evident
    when looking at the evolution of luminosity as a function of the
    mass of the core (see Fig. \ref{fig:McL}). Within this picture,
    the impact of the chemical discontinuity happens as the core grows
    and the burning shell approaches the discontinuity. Then, it
    is not completely correct to neglect the increase in the core mass
    as the luminosity drops.  From B to C the core mass changes by
    $\delta M_c /M_c =0.027366$.  Recalling that for $\nu=13$
    (eqs. \ref{eq:sh-thomson-cno_1} to \ref{eq:sh-thomson-cno_4}) now $L_s\propto {M_c}^7\mu_{\rm
      eff}^7$ this decreases the shell homology prediction of the
    luminosity drop to $\delta L/L =-0.3804$. This is still a factor
    2.8 difference from that observed in FEMs (a $\sim 180$\%
    difference in $\delta L_s/L_s$), but a significant improvement
    over the naive estimation.

  We show in the next section that the final key missing ingredient
  comes from the feedback of the changes in the shell temperature on
  the inner mantle (the isothermal mantle below the burning shell).

\subsection{The feedback on the inner mantle}
\label{sec:feedback}
As summarized in the appendix \ref{app:shellhomology}, shell-source homology relations predict that
the luminosity of the burning shell depends on $M_c$, $R_s$, and $\mu$.
For a typical Thomson scattering opacity law, and $\nu=13$, this dependence is as
\begin{equation}
L_s\propto {M_c}^7 {R_s}^{-16/3} \mu_{\rm env}^7,
%\label{eq:SH_L}
\end{equation}
or
\begin{equation}
\frac{\delta L_s}{L_s}= 7\frac{\delta M_c}{M_c}-\frac{16}{3} \frac{\delta R_s}{R_s}+7\frac{\delta\mu_{\rm env}}{\mu_{\rm env}}.
\label{eq:SH_L}
\end{equation}

Interestingly, as clarified by eqs. \ref{eq:1} and \ref{eq:2} a change
in the mean molecular weight of the outer mantle ($\mu_{\rm env}$), in a model of given core mass ($M_c$), will affect the radius
of the burning shell ($R_s$). This is because, any drop (increase) in the
temperature of the shell leads to a drop (increase) in the temperature
of the ideal gas layers immediately below the burning shell, with the
consequent contraction (expansion) of those layers. Interestingly, a
drop in the radius of the burning shell will lead to higher
temperatures than if the radius were to stay fixed.

We can determine from eqs.  \ref{eq:1} and \ref{eq:2} how changes
in $\mu_{\rm env}$ and $R_s$ are connected for the typical core masses at which
the RGBB takes place. Assuming, as before, that $\mu_s\propto \mu_{\rm env}$, eq.  \ref{eq:2} can be written as
\begin{equation}
  \begin{aligned} 
\mathcal{C}\simeq &{T_9}^{-1/6} \left[\frac{\mu_{\rm env}}{\mu_c}\right]^3 \left[\frac{R_s}{R_{dc}}\right]^2\left[\frac{M_c}{M_\odot}\right]^{-2/3} \\ 
 \times & \exp\left[10.4\frac{\mu_c}{\mu_{\rm env}}\left(1-\frac{R_{\rm s}}{R_{dc}}\right) \right.
   -\left.15.231 {T_9}^{-1/3}\right],
 \end{aligned} 
 \label{eq:3} 
\end{equation}  
 where we have used that $\nu=13$ as before. Calling
  $x=R_s/R_{\rm dc}$, $z=\mu_{\rm env}/\mu_c$ and $m=M_c/M_\odot$ it
  is easy to show from eqs. \ref{eq:1} and \ref{eq:3} that, for a
  constant mass of the core\footnote{This means that also $R_{\rm dc}$
    can be considered constant.} we have
\begin{equation}
 \begin{aligned}
   0\simeq &\frac{17}{6}\frac{\delta z}{z}+\frac{13}{6}\frac{\delta x}{x}- 10.4(1-x)\frac{\delta z}{z^2}-10.4\frac{\delta x}{z}\\
   - &(18.058\, m^{-4/9})\times\left[\frac{z^{-1/3}x^{-2/3}}{3}\delta x -\frac{z^{-4/3}x^{1/3}}{3}\delta z\right].
 \end{aligned}     
\end{equation}  
Replacing the typical values of $m$, $x$ and $z$ for the stellar structure near the RGBB, $m\approx 0.2384$, $x\approx 1.495$, and $z\approx 0.53$
we find that
\begin{equation}
  \frac{\delta R_s}{R_s}\simeq 0.66 \frac{\delta \mu_{\rm env}}{\mu_{\rm env}}.
  \label{eq:feedback}
\end{equation}
Had we assumed that $\mu_s$ remained unchanged while $\mu_{\rm env}$ changed, then the proportionality constant in eq. \ref{eq:feedback} would have been 0.64. Similarly, had we assumed a value of $\nu=16$, the constant in   eq. \ref{eq:feedback} would have remained basically unchanged at $0.66$. 
Eq. \ref{eq:feedback} is key to understand the luminosity drop at the
RGBB. Eq.  \ref{eq:feedback} tells us that a drop in the mean
molecular weight will create a similar drop in the radius of the
burning shell. This $\delta R_s$ will act to increase the temperature
and attenuate the impact of $\delta \mu_{\rm env}$ on the luminosity
of the burning shell. As discussed in Section \ref{sec:jcd}, when
  there is a chemical discontinuity in the outer mantle, $\mu_{\rm env}$ in
  the previous equations must be replaced by $\mu_{\rm eff}$ (eq. \ref{eq:mueff}).

If we use the result from eq. \ref{eq:feedback} in eq. \ref{eq:SH_L}
we see that 
\begin{equation}
\frac{\delta L_s}{L_s}\simeq 7\frac{\delta M_c}{M_c}+3.48 \frac{\delta\mu_{\rm eff}}{\mu_{\rm eff}},
\label{eq:SH_feedback}
\end{equation}
where it becomes clear how the feedback in the radius of the burning
shell effectively decreases the impact of the change in the mean
molecular weight.

If we now we use eq. \ref{eq:SH_feedback} to assess the drop in
luminosity from model B to model C ($\delta M_c/M_c=0.027366$ and
$\delta \mu_{\rm eff}/\mu_{\rm eff}=-0.0818$), we get $\delta
  L/L\simeq -0.093$ which is only 32\% less than the actual value
  observed in the FEM. This small difference
  is a huge improvement over the differences obtained in the previous
  sections, when the feedback of the core was neglected.

We see that, when the feedback of the inner mantle is included in our
estimations of the luminosity drop, shell-source homology relations do a
remarkable job in explaining the observed luminosity drop at the RGBB.
Moreover, we can now use these results to understand why the
luminosity increase slows down from model A to model B (see
Fig. \ref{fig:McL}). From Table \ref{tab:snapshots} we see that, from
model A to model B the core increases by $\delta M_c/M_c\simeq
0.04$. In the absence of any other effect this would translate into an
increase of the shell luminosity of $\delta L/L\simeq 0.28$, which is
similar to the value between A and $B^*$ in Fig. \ref{fig:McL}, but
significantly higher than the increase in luminosity between A and B
(see Fig. \ref{fig:McL}). However, when we take into account that the
effective mean molecular weight ($\mu_{\rm eff}$, eq. \ref{eq:mueff})
decreases as the burning shell gets closer to the discontinuity by
$\delta \mu_{\rm eff}/\mu_{\rm eff}\simeq -0.033$ we see that the
expected change in luminosity should be $\delta L/L\simeq
  0.165$, which is, within the quoted precision, equal to the actual
  luminosity change in the FEM.

We conclude that, when the feedback of changes in the shell
  temperature are included in the inner isothermal mantle and,
  consequently, in the location of the shell, shell-source homology relations
  do a remarkable job at explaining the luminosity changes observed on
  the RGBB.

\section{A simple description of the RGBB}
\label{sec:toy}
In the previous section we have shown that, by taking into account the
feedback of shell temperature changes into the inner mantle and
consequently the location ($R_s$) of the burning shell, shell-source homology
relations are able to quantitatively explain in luminosity around the
RGBB.  For the sake of completeness in this section we show that a
very simple description of the RGBB, useful for pedagogical purposes,
can be constructed with minimal assumptions.

\begin{figure}
\centering \includegraphics[width=\columnwidth]{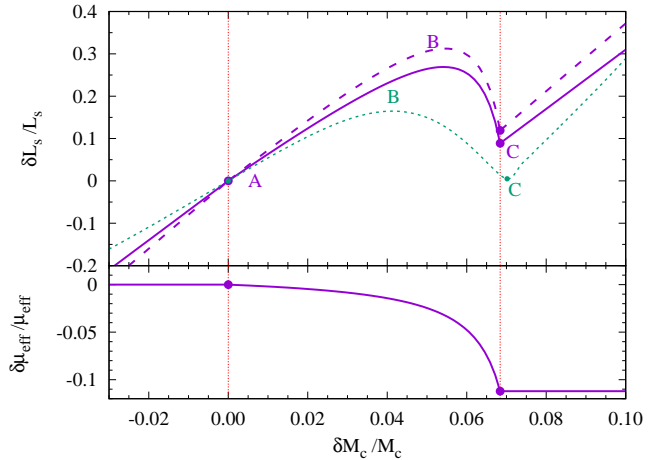}
\caption{Upper panel: The solid curve shows the relative changes
    in the shell luminosity as a function of the relative changes in
    core mass according to eq. \ref{eq:toy_L} ($\nu=13$). The dashed
    purple curve shows relative changes in the shell luminosity when
    $\nu=16$ is adopted in the derivation of shell-source homology
    relations. The green dotted curve shows the evolution of the FEMs presented in Fig. \ref{fig:HR}. Bottom panel:
  Relative changes in the effective mean molecular weight assumed in
  our toy model (eq. \ref{eq:toy_mu}) as the burning approaches the
  chemical discontinuity.  Evolution to the left and to the right of
  the vertical dashed lines proceeds without changes in the mean
  molecular weight $\mu$, and according to $\delta L/L= 7\delta
  M_c/M_c$ for $\nu=13$ (solid purple line) and $\delta L/L= 8\delta
  M_c/M_c$ for $\nu=16$ (dashed purple line).}
\label{fig:toyRGBB}
\end{figure}

Under the assumption that the structure of the outer mantle does not change dramatically from A to C,
we can write that the mass between the burning shell and the chemical discontinuity ($\Delta m$) is
\begin{equation}
  \Delta m\simeq 4\pi {R_s}^2 \bar{\rho}_{UM} (R_{\rm dis}-R_s),
  \label{eq:9}
\end{equation}  
where $\bar{\rho}_{UM}$ is some mean density above the burning
shell. Calling $\delta M_c$ the increase in mass of the core since
point A, we have $\Delta m=\Delta m^0-\delta M_c$, where $\Delta
m^0={M_c}^C-{M_c}^A$. Using this in eq. \ref{eq:9} we see that, at
first order, we can write
\begin{equation}
  \frac{R_{\rm dis}}{R_s}\simeq A-B\frac{\delta M_c}{M_c}
  \label{eq:radius_ratio}
\end{equation}  
Where the values of $A$ and $B$ in our model can be derived from the
values on Table \ref{tab:snapshots}.  Using eq. \ref{eq:radius_ratio}
in eq. \ref{eq:mueff} we can derive that the change in the effective
mean molecular weight as the burning shell advances from A to C is
\begin{equation}
\frac{\delta \mu_{\rm eff}}{\mu_{\rm eff}}\simeq \alpha\times\left[1-\frac{1}{\left(1-\beta \frac{\delta M_c}{M_c}\right)}\right],
\label{eq:toy_mu}
\end{equation}  
with $\alpha=0.012936$ and $\beta=13.1048$. Eq. \ref{eq:toy_mu}
captures the essence of the change in the molecular weight as the
burning shell approaches the chemical discontinuity. Together with eq. \ref{eq:feedback}
we see that, as the burning shell advances from model A to model C, the luminosity
will follow
\begin{equation}
\begin{aligned}  
  \frac{\delta L_s}{L_s}\simeq &7\frac{\delta M_c}{M_c}+ 3.48\,  \alpha\, \left[1-\frac{1}{\left(1-\beta \delta M_c/M_c\right)}\right].
\end{aligned}  
\label{eq:toy_L}
\end{equation}  
The evolution described by eq. \ref{eq:toy_L} is shown in
Fig. \ref{fig:toyRGBB} where it is compared with the behavior of
  the FEM shown in Fig. \ref{fig:McL}. We see that
  the simple model presented in this section captures very well the
  behavior shown by FEMs. The main difference
  between the simple model and the full evolutionary structures arises
  from the fact that eq. \ref{eq:toy_mu} assumes that the mean
  molecular weight below the discontinuity has a constant value. As it
  is clear from Fig. \ref{fig:r_m_H} in FEMs the
  hydrogen profile (and consequently $\mu$) has a nonzero slope. As a
  consequence, in FEMs as the shell approaches the
  discontinuity the mean molecular weight decreases faster than in our
  toy model due to this effect. This leads to an additional decrease
  in the luminosity, leading to a slightly smaller slope in the
  $L_s(M_c)$ relationship. In fact, once the shell reaches the
  discontinuity, and the shell evolves through an homogeneous layer
  (after point C), the FEMs show a very similar
  slope to that predicted by shell-source homology relations ---in particular
  for $\nu=16$ which is the correct temperature dependence of CNO
  burning at those temperatures.

\section{Discussion and conclusions}
\label{sec:end}
We have reanalyzed the properties of the RGBB in the light of our
recent description of the properties of red giants
\citep{2022-ToyStory}.  Specifically, we have made use of a simple
description of the structure of red giants that includes the
connection between the location of the burning shell, the mean
molecular weight of the outer mantle and the mass of the core. With
the help of this simple model we have shown in Section
\ref{sec:feedback} that, when the mean molecular weight drops during
the RGBB, the feedback of the temperature of the burning shell on the
inner isothermal mantle, leads to a decrease in the radius of the
burning shell that attenuates the luminosity drop. When this
  feedback is taken into account together with the description by
  \cite{2015MNRAS.453..666C}, of how the effective mean molecular
  weight of the outer mantle changes as the burning shell approaches
  the chemical discontinuity, shell-source homology relations are
completely able to quantitatively explain what is observed in FEMs. Specifically, when taking into account the
  increase of the core mass and the decrese in the effective mean
  molecular weight as the burning shell approaches the discontinuity,
  together with feedback of the inner mantle the predictions of shell
  homology relations are in agreement with FEMs. This definitely clarifies the role played by each part of
  the star in the formation of the RGBB and how the luminosity changes
  as the burning shell approaches the chemical disconinuity.

Moreover, in Section \ref{sec:toy-model}, we have shown that
  the theoretical framework developed in \cite{2022-ToyStory} can be
  used to give a more specific meaning to the quantities involved in
  shell-source homology relations. In particular, this approach
  clarifies which value of $\mu$ (i.e. at which point in the star) is
  relevant for shell-source homology relations.

In addition, we have shown in Section \ref{sec:toy} that the whole
evolution of the stellar luminosity before and after the RGBB can be
described with a simple model that takes into account the previously
mentioned feedback and the change in the effective mean molecular
weight as the burning shell approaches the discontinuity. Most
importantly, this description emphasizes that both the initial slowing
down of the luminosity increase (from model A to B, Figs. \ref{fig:HR}
and \ref{fig:McL}), and the posterior drop in luminosity (from model B
to C, Figs. \ref{fig:HR} and \ref{fig:McL}) all take place on a
nuclear timescale (i.e. in thermal equilibrium), as the burning shell
burns its way towards the chemical discontinuity left by convection.
Consequently, this toy model demonstrates that the luminosity
  changes during the RGBB are just a consequence of the changes in the
  temperature of the burning shell produced by variations of the
  effective mean molecular weight of the outer mantle, and the
  consequent impact of those temperature changes in structure of the
  inner mantle of the red giant (Fig. \ref{fig:Structure}). We believe
  this toy model has great pedagogical potential for discussing the
  RGBB. 

In closing we would like to mention that the clear description of the
RGBB obtained here, from the simple models devised in
\cite{2015MNRAS.453..666C} and \cite{2022-ToyStory}, highlights the
importance of simple mental models when interpreting and understanding
the results from detailed numerical simulations.

%--------------EJEMPLO

\acknowledgments The author thanks the Max Planck Institute for
Astrophysics and Achim Weiß for several research stays during which
many of the ideas in this work were conceived and developed. The
author also thanks Jørgen Christensen-Dalsgaard, Alfred Gautschy, and
the anonymous referee for comments and corrections that highlighted
shortcomings in the first version of this paper. M3B is partially
supported by PIP 2971 from CONICET and PICT 2020-03316 from Agencia
I+D+i.


\vspace{5mm}
%%\facilities{HST(STIS), Swift(XRT and UVOT), AAVSO, CTIO:1.3m, CTIO:1.5m,CXO}


\software{{\tt LPCODE:} \cite{2003A&A...404..593A,2005A&A...435..631A,2016A&A...588A..25M,2020A&A...644A..55A}}



\appendix

\section{Shell-source homology relations with varying $\mu$}
\label{app:shellhomology}
      Shell-source homology was first introduced by \cite{1970A&A.....6..426R} and is based on several simplifying assumptions:
        \begin{enumerate}
\item It is assumed that there exist a region of negligible mass ($\Delta
m\ll M_c$) from the bottom of the burning shell ($r=r_{-}$) to a point
$r=R_0$ above the burning shell ($R_0>r_{+}$), where $T$, $P$, and
$\rho$ decrease significantly from their values at the burning shell
(i.e. $T(R_0)\ll T_s$, $P(R_0)\ll P_s$, $\rho(R_0)\ll \rho_s$, and $l(R_0)=l(r_{+})=L_s+L_c$, where usually $L_c=0$). 
\item That region of the star is assumed in ``thermal equilibrium'' (i.e. $dl/dm= \epsilon_n$).
\item The gas is considered to be an ideal classical gas $P=\Re \rho
  T/\mu $. Note that \cite{1970A&A.....6..426R} extends this to a
  classical gas plus radiation.
\item Heat is transported by radiation in the whole region.
\item Physical quantities in the region of concern ($r_{-}<r<R_0$) are
  only sensitive to the radius ($R_s$) and mass of the core ($M_c$),
  and to a characteristic mean molecular weight of the material in the
  region ($\bar{\mu}$), in the sense that a set of solutions of the
  stellar structure equations $\rho(r)$, $T(r)$, $P(r)$, $l(r)$
  (corresponding to $M_c$,$R_s$,$\bar{\mu}$) and a  set of solutions of the
  stellar structure equations $\rho'(r')$, $T'(r')$, $P'(r')$, $l'(r')$
  (corresponding to $M_c'$,$R_s'$,$\bar{\mu}'$) evaluated at
  homologous points ($r/R_s=r'/R_s'$) are related by
%\begin{equation}
  \begin{align}
    \frac{\rho}{\rho'}=& \left(\frac{M_c}{M_c'}\right)^{\varphi_1} \left(\frac{R_s}{R_s'}\right)^{\varphi_2}\left(\frac{\bar{\mu}}{\bar{\mu}'}\right)^{\varphi_3}\label{SH_1}\\
    \frac{T}{T'}=&\left(\frac{M_c}{M_c'}\right)^{\psi_1} \left(\frac{R_s}{R_s'}\right)^{\psi_2}\left(\frac{\bar{\mu}}{\bar{\mu}'}\right)^{\psi_3}\label{SH_2}\\
    \frac{P}{P'}=& \left(\frac{M_c}{M_c'}\right)^{\tau_1} \left(\frac{R_s}{R_s'}\right)^{\tau_2}\left(\frac{\bar{\mu}}{\bar{\mu}'}\right)^{\tau_3}\label{SH_3}\\
    \frac{l}{l'}=& \left(\frac{M_c}{M_c'}\right)^{\sigma_1} \left(\frac{R_s}{R_s'}\right)^{\sigma_2}\left(\frac{\bar{\mu}}{\bar{\mu}'}\right)^{\sigma_3},
    \label{SH_4}
 \end{align}
%\end{equation}  
        \end{enumerate}          
        When working with shell-source homology relations it is also typical to assume that the massive envelope is also in thermal equilibrium and $L_s=L_\star$, although this is not needed to derive the behavior of the burning shell but to link it to the surface luminosity of the star.

Assuming power laws for the specific energy generation rate $\epsilon$ and the radiative opacity
$\kappa$ ($\epsilon\propto \rho^{n-1} T^\nu$ and $\kappa\propto
  P^a T^b$) it is possible to show that the coefficients in eqs. \ref{SH_1} to \ref{SH_4}  fulfill.
\begin{equation}
  \begin{aligned}
    \psi_1=&1,\ \ \ \ \ \ \varphi_1=\frac{4-\nu-a-b}{1+a+n},\\
    \tau_1=&\varphi_1+1,\ \ \sigma_1=\varphi_1 n + \nu,\\
    \psi_2=&-1,\ \ \ \ \varphi_2=\frac{-6+\nu+a+b}{1+a+n},\\
    \tau_2=&\varphi_2-1,\ \ \sigma_2=\varphi_2 n - \nu+3,\\
    \psi_3=&1,\ \ \ \ \ \ \varphi_3=\frac{4-\nu-a-b}{1+a+n},\\
    \tau_3=&\varphi_3,\ \ \sigma_3=\varphi_3 n + \nu. 
    \label{SH-coef}
 \end{aligned}
\end{equation}        
 A detailed explanation of how to obtain these results can be found in
 Chapters §33.2 and §33.3 of \cite{2012sse..book.....K}, and in the
 original article by \cite{1970A&A.....6..426R}. As discussed in
 section \ref{sec:toy-model} the meaning of $\bar{\mu}$ is not well
 defined. The most natural way by which one can characterize the chemical
 composition of the relevant region by only one parameter
 ($\bar{\mu}$) is by assuming that the function $\mu(r/R_s)$ is
 related to $\mu'(r'/R_s')$ by a single factor
 $\mu(r/R_s)/\mu'(r'/R_s')= \bar{\mu}/\bar{\mu}'$. This is not
 realistic as, at the bottom of the burning shell of two stellar
 models the mean molecular weight must be that of the core
 $\mu(r_{-}/R_s)=\mu'(r_{-}'/R_s')=\mu_c$. Alternatively, one can
 assume $\bar{\mu}$ to represent some ill-defined value of the mean
 molecular weight of the whole region. As discussed in section
 \ref{sec:toy-model}, the framework developed by \cite{2022-ToyStory}
 shows that $\bar{\mu}$ is very close to the mean molecular
 weight immediately above the burning shell ($\bar{\mu}\simeq \mu_{\rm
   env}$) with very minor corrections (see
 eqs. \ref{eq:mb22-thomson-cno_1} to \ref{eq:mb22-thomson-cno_4}).

For the conditions in red giants, the values of $a=b=0$ (Thomson
scattering) and $\nu=13$, $n=2$ have been used extensively
\citep{2012sse..book.....K}. This implies that the dependence of the
shell luminosity is $L\propto {M_c}^7 {R_c}^{-16/3}
\bar{\mu}^7$. Thomson scattering is, in fact, a very good
  approximation of  the conditions at the burning shell in red
  giants.  At the typical lower temperatures of the RGBB ($T_s\simeq
2.85 \times 10^7$K) the temperature dependence of the CNO cycle is
closer to $\nu=16$. With these choices the dependence of the shell
luminosity is $L\propto {M_c}^8 {R_c}^{-19/3} \bar{\mu}^8$, which
gives results closer to the predictions of FEMs.
\bigskip

\bibliography{biblio}{}
\bibliographystyle{aasjournal}



\end{document}